\documentstyle[12pt]{article}
\title{A new field--theoretical formulation for the motion of an
electron in a quenched disorder potential
\thanks{Contribution to the 5th International Conference on Path
Integrals, Dubna 1996}}
\author{\large W. Weller,
F. Stefani\thanks{Institut f\"ur Angewandte Geod\"asie, Leipzig} ,
M. Souleiman\thanks{ZMAI, Medizinische Fakult\"at, Universit\"at
Leipzig} \\[3mm]
\em Fakult\"at f\"ur Physik und Geowissenschaften,  \\
\em Universit\"at Leipzig, \\
\em Augustusplatz, D 04109 Leipzig, Germany}
\date{}

\input prepictex
\input pictex
\input postpictex

\begin{document}
\maketitle
\begin{abstract}
Following a proposal by Aronov and Ioselevich, we express the
Green functions (GF) of a noninteracting disordered Fermi system
as a functional integral on a real time/frequency lattice. The
normalizing denominator of this functional integral is equal to unity,
because of identities satisfied by the GF.
The GF can then be simply averaged with respect to the
random disorder potential.
We describe the
fermionic fields not belonging to the external
frequency by means of a bosonic auxiliary field $g$. The
Hubbard--Stratonovich field $Q$ is introduced only with respect to
the fermionic fields for the external frequency.
\end{abstract}

\section{Introduction}

We consider noninteracting Fermions (electrons) under the influence of
a random quenched disorder potential.
The Green functions (GF) or products of GF are usually
expressed as functional integrals with a normalizing denominator.
It is this normalizing denominator which prevents simple averaging
with respect to a model--like Gaussian distributed random potential.
The normalizing denominator is
equal to unity in the replica method \cite{Edwards},
the supersymmetry method \cite{Efetov},
the Keldysh method \cite{Babichenko},
and a method proposed by Aronov and Ioselevich
\cite{Aronov}, using one real time axis.
\par

In this work we are concerned with further development of the method
of Aronov and Ioselevich. In Section 2, we start like these
authors with a  definition of the GF on a time lattice, what
allows definition of the functional integral as a multiple integral.
In frequency representation the definition of the functional
integral needs a frequency lattice; we call the method
``Many--frequency--technique''.
We define here the GF as periodic in the time difference
by adding a boundary term in the definition. From the
definition of the GF follow well defined
identities for the GF (Eqs. (\ref{26}), (\ref{26b}) below),
necessary for the vanishing of all closed loops.
In order to avoid complications due to so many
frequencies, we treat in Section 2.1
the fermionic fields not belonging to the external frequency
by means of one bosonic auxiliary field $g$.
A differentiation ``mechanism'' with respect to
$g \rightarrow 0$ takes care of eliminating the contribution
of the closed loops.
Finally, in Section 2.2, we introduce the Hubbard--Stratonovich field
$Q$ only for the fermionic fields referring to the external frequency.
\par

\section{Many--frequency--technique}

On the real time interval $[-T, T)$ a time lattice is introduced by
\begin{equation}
\label{21}
\Delta = \frac{T}{{\cal N}} \;,\;\;\; t_n = n\Delta \;, \;
n = -{\cal N}, -{\cal N}+1, \ldots, {\cal N} -1\;.
\end{equation}
We define the real time retarded and advanced GF corresponding to
an eigenvalue $E_{\lambda}$ for the one--particle system by
\begin{eqnarray}
\label{22}
&&(G_{R \lambda})^{-1} (t_{n_1} - t_{n_2}) = i \delta_{n_1, n_2}
- i b_{R \lambda}
\big[ \delta_{n_1-1, n_2}(1 - \delta_{n_1, -{\cal N}}) +
\delta_{n_1, -{\cal N}}\delta_{{\cal N}-1, n_2} \big]  \; ,
\nonumber \\
&&(G_{A \lambda})^{-1} (t_{n_1} - t_{n_2}) = -i \delta_{n_1, n_2}
+ i b_{A \lambda}
\big[ \delta_{n_1+1, n_2}(1 - \delta_{n_1, {\cal N}-1}) +
\delta_{n_1, {\cal N}-1} \delta_{-{\cal N}, n_2} \big]  \; ,
\nonumber \\
&&b_{R, A \lambda} = \exp \big[\mp i(E_{\lambda} - E_F \mp i \eta) \Delta \big]
\approx 1 \mp i(E_{\lambda} - E_F \mp i \eta) \Delta   \; ;
\end{eqnarray}
$E_F$ is the Fermi energy.
The definition (\ref{22}) coincides
with that of Aronov and Joselevich \cite{Aronov} besides the added
boundary terms.
The following procedure for the limits is used:
\begin{equation}
\label{23}
\mbox{first} \;\;\;  {\cal N} \rightarrow \infty  \;\;\;
\mbox{with} \;\;\; {\cal N} \Delta = T = \mbox{const} \; , \;
\mbox{then} \;\;\; T \rightarrow \infty \; ,  \;
\mbox{finally} \;\;\; \eta \rightarrow +0 \; .
\end{equation}
The Fourier transform of the GF is
\begin{eqnarray}
\label{24}
&&G_{\alpha \lambda}(\omega_l)
= \frac{1}{\Delta} G_{\alpha \lambda}^{cont}(\omega_l)
= \frac{-i\alpha}{1 - \exp \big\{i\alpha(\omega_l - E_{\lambda} + E_F
+ i\alpha \eta)\Delta \big\}} \\
&&\approx
\frac{1}{\Delta} \; \frac{1}{\omega_l - E_{\lambda} + E_F + i\alpha \eta}
\;\;\; \mbox{for} \;\;\; |\omega_l \Delta| \ll 1  \; .  \nonumber
\end{eqnarray}
Here $\alpha = R, A \;\; \mbox{or} +1, -1 \; ; \;
\omega_l = (2\pi/2T) l \;\; \mbox{with} \;\;
l = -{\cal N}, -{\cal N}+1, \ldots, {\cal N}-1 \; $.
Expression (\ref{24}) shows  the
coincidence of the lattice GF
with the continuum GF in the limit $\Delta \rightarrow 0$.
\par

The GF are expressed (in unperturbed basis)
by the Gra{\ss}mann functional integral
\begin{eqnarray}
\label{25}
&&G_{\alpha; \lambda_1 \lambda_2}(\omega_l) =
\int {\cal D}\psi ^+{\cal D}\psi \;\; \psi_{\alpha l \lambda_1} \;
\psi^+_{\alpha l \lambda_2}
\exp(- S) \;,  \\
&&S=\sum_{i l \lambda_1 \lambda_2}
\psi^+_{\alpha_i l \lambda_1}
\big( G_{\alpha_i}
(\omega_l) \big)^{-1}_{\lambda_1 \lambda_2}
\psi_{\alpha_i l \lambda_2}
\;,  \nonumber   \\
&&\big( G_{\alpha} (\omega_l) \big)^{-1}_{\lambda_1 \lambda_2} =
(\lambda_1|
i \alpha \Big[ 1 - \exp \Big\{ i\alpha \big( \omega_l
- h^{(0)} + E_F + i\alpha \eta - v({\bf r})
\big) \Delta \Big\} \Big] |\lambda_2)
\nonumber \\
&&\approx (\lambda_1|
i \alpha \Big[
1 - \exp \{ i\alpha \omega_l \Delta \}
\Big\{ 1 + i\alpha \big( h^{(0)} - E_F - i\alpha \eta + v({\bf r})
\big) \Delta \Big\}
\Big]
|\lambda_2) \; .  \nonumber
\end{eqnarray}
Here, instead of the eigenvalue $E_{\lambda}$,
the one--particle Hamiltonian
$h^{(0)} - E_F + v({\bf r})$ is used with $h^{(0)}$ as an
unperturbed Hamiltonian with eigenstates $|\lambda)$ and
eigenvalues $E^{(0)}_{\lambda}$;
$v({\bf r})$ is the random potential.
Because averages of the product of two GF are needed, two Gra{\ss}mann
fields are introduced; the $\alpha_i \; (i = 1, 2)$ are
$R, R$ or $A, A$ or $R, A$, respectively.    \par

Eq. (\ref{25}) contains {\it no} normalizing denominator;
it is an identity because of the
identity (written in diagonal representation)
for the GF (see \cite{Aronov}):
\begin{equation}
\label{26}
Z: =
\int {\cal D}\psi ^+{\cal D}\psi \; \exp(- S)
= \prod_{\alpha \lambda}
\det \Big[ \big( G_{\alpha \lambda} \big)^{-1}
(t_{n_1} - t_{n_2}) \Big] =
\prod_{\alpha l \lambda} (G_{\alpha \lambda})^{-1} (\omega_l) = 1 \;.
\end{equation}
The matrix
$\big( G_{R \lambda} \big)^{-1}(n_1 - n_2)$ in $n_1, n_2$ has only
elements for $n_2=n_1$, for $n_2=n_1-1$ and a single element for
$n_1=-{\cal N}, n_2={\cal N}-1$. Expansion of the determinant with
respect to the elements of the first row gives
in accordance with the procedure (\ref{23}) for the limits:
\begin{equation}
\label{26a}
\det \Big[ \big( G_{R \lambda} \big)^{-1} \Big] =
1 + i b_{R \lambda} (- i b_{R \lambda})^{2{\cal N}-1} =
1 - \exp \big[- i(E_{\lambda} - i \eta) (2{\cal N}-1) \Delta \big]
\rightarrow 1 \; .
\end{equation}
\par
From (\ref{26}) follows the identity
\begin{equation}
\label{26b}
\sum_l \Delta \exp \{i\alpha \omega_l \Delta\}
G_{\alpha \lambda} (\omega_l) = 0
\end{equation}
by differentiation with respect to $E_{\lambda}$;
similar identities hold for $l$--sums over products of {\it only}
retarded or {\it only} advanced GF in any basis.
In a perturbation expansion these identities
guaranty the vanishing of all closed loops,
because a closed loop leads to a
frequency sum.   \par

Now, the absence of a normalizing denominator in (\ref{25})
allows simple averaging (denoted by $<\ldots >_v$) over the Gaussian
distributed random potential:
\begin{eqnarray}
\label{29}
&&<G_{\alpha; \lambda_1 \lambda_2}(\omega_l)>_v =:
<\psi_{\alpha l \lambda_1} \; \psi^+_{\alpha l \lambda_2}>_v =
\int {\cal D}\psi ^+{\cal D}\psi \;\; \psi_{\alpha l \lambda_1} \;
\psi^+_{\alpha l \lambda_2}
\exp(- S_{eff}) \;,  \nonumber \\
\label{29a}
&&S_{eff} =\sum_{i l \lambda}
\psi^+_{\alpha_i l \lambda}
\big( G^{(0)}_{\alpha_i \lambda}
(\omega_l)\big)^{-1}
\psi_{\alpha_i l \lambda}  \\
&&- \frac{\gamma \Delta^2}{2}
\sum_{i l \lambda_1 \lambda_2 ;
j \bar{l} \bar{\lambda}_1 \bar{\lambda}_2}
\Gamma^{\lambda_1 \lambda_2}
_{\bar{\lambda}_1 \bar{\lambda}_2}
\exp \{ i\alpha_i \omega_l \Delta \}
\psi^+_{\alpha_i l \lambda_1}
\psi_{\alpha_i l \lambda_2}
\exp \{ i\alpha_j \omega_{\bar{l}} \Delta \}
\psi^+_{\alpha_j \bar{l} \bar{\lambda}_1}
\psi_{\alpha_j \bar{l} \bar{\lambda}_2} \;,  \nonumber \\
&&\gamma \Gamma^{\lambda_1 \lambda_2}
_{\bar{\lambda}_1 \bar{\lambda}_2} =
(\lambda_1| (\bar{\lambda}_1|
<v({\bf r}) v(\bar{{\bf r}})>_v
|\bar{\lambda}_2) |\lambda_2) \; ;
\end{eqnarray}
$\gamma$ is the strength of the potential correlator.
\par

\subsection{Introduction of the ${\bf g}$--field}

Besides the retarded and the advanced fields for a fixed external
frequency $\omega_{l_0}$, all fields belonging to other frequencies
shall be treated ``globally'' by a bosonic auxiliary field $g$. For
this, we introduce into the functional integral the identities
\begin{equation}
\label{32}
1 = \int_{-\infty}^{\infty} dg_{\lambda_1 \lambda_2}
\delta \Big\{ g_{\lambda_1 \lambda_2} - \sum_{i l; (l\not= l_0)}
\exp \{ i \alpha_i \omega_l \Delta \}
\psi^+_{\alpha_i l \lambda_1} \psi_{\alpha_i l \lambda_2}  \Big\} \; .
\end{equation}
This changes the measure in the functional integral
to $\exp\{-S_{\psi}\} :=$
\begin{eqnarray*}
\label{33}
= \exp \Big[- \sum_{i \lambda_1 \lambda_2 l ; (l \not= l_0)}
\psi_{\alpha_i l \lambda_1}^+
\Big\{ \big( G^{(0)}_{\alpha_i \lambda_1} (\omega_l) \big)^{-1}
\delta_{\lambda_1 \lambda_2}
- \sqrt{\gamma} \Delta \exp\{i\alpha_i \omega_l\Delta\}
\frac{\partial}{\partial g_{\lambda_1 \lambda_2}} \Big\}
\psi_{\alpha_i l \lambda_2}
\Big] \ast  &&   \\
\exp \Bigg[\frac{1}{2} \sum_{
\lambda_1 \lambda_2; \bar{\lambda}_1 \bar{\lambda}_2}
\Gamma^{\lambda_1 \lambda_2} _{\bar{\lambda}_1
\bar{\lambda}_2}
g_{\lambda_1 \lambda_2}
g_{\bar{\lambda}_1 \bar{\lambda}_2}
- \sum_{i \lambda_1 \lambda_2}
\psi_{\alpha_i \lambda_1}^+
\Big\{ \big( G^{(0)}_{\alpha_i \lambda_1} \big)^{-1}
\delta_{\lambda_1 \lambda_2}
- \sqrt{\gamma} \Delta \ast \hspace*{1.7cm}  (\arabic{equation})
&& \\
\sum_{\bar{\lambda}_1 \bar{\lambda}_2}
\Gamma^{\lambda_1 \lambda_2}_{\bar{\lambda}_1
\bar{\lambda}_2}
g_{\bar{\lambda}_1 \bar{\lambda}_2}
\Big\} \psi_{\alpha_i \lambda_2}
+\frac{\gamma \Delta^2}{2}
\sum_{\lambda_1 \lambda_2; \bar{\lambda}_1
\bar{\lambda}_2}
\Gamma^{\lambda_1 \lambda_2}_{\bar{\lambda}_1
\bar{\lambda}_2}
\Big\{ \sum_i  \psi^+_{\alpha_i \lambda_1}
\psi_{\alpha_i \lambda_2} \Big\}
\Big\{ \sum_j  \psi^+_{\alpha_j \bar{\lambda}_1}
\psi_{\alpha_j \bar{\lambda}_2} \Big\}
\Bigg] \Bigg|_{g=0} .        &&
\end{eqnarray*}
In obtaining (\ref{33}) we used the integral representation of the
$\delta$--function (with a field $\vartheta_{\lambda_1 \lambda_2}$ in the
exponent), and rescaled the integration variables,
$\sqrt{\gamma} \Delta g = \tilde{g} \;\; , \;\;
\vartheta/\sqrt{\gamma} \Delta$
\linebreak[2] $= \tilde{\vartheta}$ (the
tilde is omitted). Further,
for $i \vartheta_{\lambda_1 \lambda_2}$ we substituted
$\partial/\partial g_{\lambda_1 \lambda_2}$ acting on
$\exp\{i \sum_{\lambda_1 \lambda_2} \vartheta_{\lambda_1 \lambda_2}
g_{\lambda_1 \lambda_2} \}$; after partial integration,
$g_{\lambda_1 \lambda_2} = 0$ results because of
$\delta$--functions. The $\psi_{\alpha \lambda}$ refer to the
frequency $\omega_{l_0}$.
Now, the $\psi$ with frequencies with $l \not= l_0$ appear only bilinearly
in the first line of (\ref{33}) and can be integrated out.
\par

\addtocounter{equation}{1}
\subsection{Introduction of the ${\bf Q}$--Field}

We introduce the $Q$--field by first introducing the
identity for the Hermitean field $\bar{Q}$,
\begin{equation}
\label{42}
1 = \int {\cal D} \bar{Q} \exp \Big\{ -\frac{1}{2\gamma}
Tr\{\bar{Q}^2\} \Big\} \; ,   \\
\end{equation}
into the functional integral, and by applying a shift,
\begin{equation}
\label{43}
\bar{Q}_{\alpha_1 \lambda_1; \alpha_2 \lambda_2} =
Q_{\alpha_1 \lambda_1; \alpha_2 \lambda_2}
+ i \gamma \Delta
\sum_{\bar{\lambda}_1 \bar{\lambda}_2}
\big( \Gamma^{1/2} \big)^{\lambda_1 \bar{\lambda}_1}
_{\bar{\lambda}_2 \lambda_2}
\psi^+_{\alpha_2 \bar{\lambda}_2} \psi_{\alpha_1 \bar{\lambda}_1}
\; ,
\end{equation}
cancelling the fourth order term in the $\psi$ in (\ref{33});
the composition rule for the $\Gamma$,
\begin{equation}
\label{44}
\sum_{\bar{\lambda}_1 \bar{\lambda}_2} \big( \Gamma^s \big)
^{\lambda_1 \bar{\lambda}_1}_{\bar{\lambda}_2 \lambda_2'}
\big(\Gamma^t \big)^{\bar{\lambda}_1 \lambda_1'}
_{\lambda_2 \bar{\lambda}_2} =
\big(\Gamma^{s+t} \big)^{\lambda_1 \lambda_1'}
_{\lambda_2 \lambda_2'} \; ,
\end{equation}
was used. By integrating out the
remaining $\psi_{\alpha \lambda}$ we obtain finally for the measure of the
functional integral $\exp \{-S_{Q}\} :=$
\begin{eqnarray*}
\label{45}
= \exp \Big[ Tr \sum_{l ; (l \not= l_0)} \ln \Big\{
\big( G^{(0)}_{\alpha_i \lambda_1} (\omega_l) \big)^{-1}
\delta_{\lambda_1 \lambda_2}
-  \sqrt{\gamma} \Delta \exp\{i\alpha_i \omega_l\Delta\}
\frac{\partial}{\partial g_{\lambda_1 \lambda_2}}
\Big\} \Big] \ast
\hspace{1.3cm}
(\arabic{equation})  && \\
\exp \Bigg[ \frac{1}{2} \sum_{
\lambda_1 \lambda_2; \bar{\lambda}_1 \bar{\lambda}_2}
\Gamma^{\lambda_1 \lambda_2}_{\bar{\lambda}_1
\bar{\lambda}_2}
g_{\lambda_1 \lambda_2}
g_{\bar{\lambda}_1 \bar{\lambda}_2}
-\frac{1}{2\gamma}
\sum_{i \lambda_1; j \lambda_2}
|Q_{\alpha_i \lambda_1; \alpha_j \lambda_2}|^2
+ Tr \ln \Big\{ \big( G^{(0)}_{\alpha_i \lambda_1} \big)^{-1}
\delta_{i j} \delta_{\lambda_1 \lambda_2}   && \\
- \sqrt{\gamma} \Delta \sum_{\bar{\lambda}_1 \bar{\lambda}_2}
\Gamma^{\lambda_1 \lambda_2}_{\bar{\lambda}_1
\bar{\lambda}_2}
g_{\bar{\lambda}_1 \bar{\lambda}_2}
\delta_{i j}
+ i \Delta \sum_{\bar{\lambda}_1 \bar{\lambda}_2}
\big( \Gamma^{1/2} \big)^{\lambda_1 \bar{\lambda}_1}
_{\bar{\lambda}_2 \lambda_2}
Q_{\alpha_i \bar{\lambda}_1; \alpha_j \bar{\lambda}_2}
+ i \Delta
\hat{Q}_{\alpha_i \lambda_1; \alpha_j \lambda_2}
\Big\} \Bigg] \Bigg|_{g=0} \; ;  &&
\end{eqnarray*}
Tr is defined with respect to $i, \lambda$,
and we added a source term with $\hat{Q}$.
\par

\addtocounter{equation}{1}
\vspace{.6cm}
\noindent
{\large {\bf Appendix}}
\par

\vspace{.3cm}  \noindent
We remark that the
identity (\ref{26b}) can be also obtained by evaluating directly
the sum over $l$ by means of the residuum theorem:
\begin{eqnarray}
\label{A3}
&&\sum_{l=-{\cal N}}^{{\cal N}-1}
f\Big( \exp \{i\frac{2\pi}{2{\cal N}} l \}\Big) =
\oint_{C_1} \frac{dl}{\exp \{ i2\pi l\} -1}
f\Big( \exp \{i\frac{2\pi}{2{\cal N}} l \}\Big)  \\
&&= \frac{2{\cal N}}{2\pi i} \oint_{C_2} \frac{dz}{z}
\frac{1}{z^{2{\cal N}} - 1} f(z) =
- 2{\cal N} \; \mbox{Res}_{\mbox{{\small (outer domain)}}} \;
\Big[\frac{1}{z}
\frac{1}{z^{2{\cal N}} - 1} f(z) \Big]
\; .  \nonumber
\end{eqnarray}
The path $C_1$ in the complex $l$--plane surrounds the poles of the
denominator in the first line of (\ref{A3}).
Introducing $z=\exp \{i2\pi l/2{\cal N} \}$, it is
substituted by the path $C_2$ (see Fig.).
In analogy to the situation in the continuum, the lattice GF $G_R$ is
analytic outside the unit circle, and $G_A$ is analytic inside the
unit circle. \par

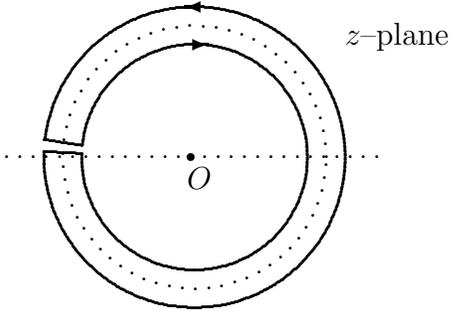
\beginpicture
\unitlength0.6cm
\setcoordinatesystem units <0.25cm,0.25cm>
\setplotarea x from 0 to 20, y from 0 to 20
\setplotsymbol ({\xpt\rm.})
\plotsymbolspacing0.5pt
\setsolid
\circulararc 355.8 degrees from 2 10.3 center at 10 10
\circulararc 355.8 degrees from 4 10.2 center at 10 10
\setlinear  \plot 2 10.93 4 10.63 /
\setlinear  \plot 2 10.3 4 10.2 /
\put {\circle*{.2}} at 10 10
\setdots
\circulararc 360 degrees from 17 10 center at 10 10
\setlinear  \plot 0 10 20 10 /
\put {$z$--plane} [Bl] at 18 16
\put {$O$} [Bl] at 9.6 8.3
\thicklines
\put {\vector(-1,0){.3}} [Bl] at 10.2 18
\put {\vector(1,0){.3}} [Bl] at 10 16
\endpicture

\noindent
{\small The path $C_2$ in the complex $z$--plane. The unit circle
and the real axis are dotted.
The poles of $1/(z^{2{\cal N}} - 1)$ on the unit circle,
lying in the interval $-\pi \le \arg z < \pi$, are
surrounded (inner domain) by $C_2$; the poles of $f(z)$ lie in the outer
domain (containing the origin $O$ and the infinite point).
}

\end{document}